# Early star-forming galaxies and the reionization of the Universe

Brant E. Robertson*, Richard S. Ellis*, James S. Dunlop¶, Ross J. McLure¶ and Daniel P. Stark§

*Department of Astronomy, California Institute of Technology MC 249-17, Pasadena CA 91125 USA*
¶*Institute for Astronomy, University of Edinburgh, Edinburgh EH9 3HJ, UK*
§*Institute of Astronomy, University of Cambridge, Cambridge CB3 0HA, UK*

**Star forming galaxies represent a valuable tracer of cosmic history. Recent observational progress with Hubble Space Telescope has led to the discovery and study of the earliest-known galaxies corresponding to a period when the Universe was only ~800 million years old. Intense ultraviolet radiation from these early galaxies probably induced a major event in cosmic history: the reionization of intergalactic hydrogen. New techniques are being developed to understand the properties of these most distant galaxies and determine their influence on the evolution of the universe.**

The frontier in completing the physical story of cosmic history is to understand *cosmic reionization* -- the transformation of neutral hydrogen, mostly located outside galaxies in an intergalactic medium (IGM) -- into an ionized state. Neutral hydrogen first formed 370,000 years after the Big Bang and released the radiation presently observed as the cosmic microwave background (CMB)[1]. Initially devoid of sources of light, the universe then entered a period termed the 'Dark Ages'[2] until the first stars formed from overdense clouds of hydrogen gas that cooled and collapsed within early cosmic structures. Observations of distant quasars[3] demonstrate that the IGM has been highly ionized since the universe was ~1 billion years old, and the transition from a neutral medium is popularly interpreted as arising from ionizing photons with energies greater than 13.6eV (wavelength λ<91.2 nm) generated by primitive stars and galaxies[4] (Fig. 1).

Astronomers wish to confirm the connection between early galaxies and reionization because detailed studies of this period of cosmic history will reveal the physical processes that originally shaped the galaxies of various luminosities and masses we see around us today. Alternative sources of reionizing photons include material collapsing onto early black holes that power active galactic nuclei, and decaying elementary particles. Verifying that star-forming galaxies were responsible for cosmic reionization requires understanding how many energetic ultraviolet (UV) photons were produced by young stars at early times and what fraction of photons capable of ionizing hydrogen *outside* galaxies escaped without being intercepted by clouds of dust and hydrogen *within* galaxies. Astronomers desire accurate measurements of the abundance of early galaxies and the distribution of their luminosities to quantify the number of sources producing energetic photons, as well as a determination of the mixture of stars, gas, and dust in galaxies to ascertain the likelihood that the UV radiation can escape to ionize the IGM[5,6]. The Lyman α emission line, detectable using spectrographs on large ground-based telescopes, is a valuable additional diagnostic given it is easily erased by neutral gas *outside* galaxies[7-12]. Its observed strength in distant galaxies is therefore a sensitive gauge of the latest time when reionization was completed.

In this primarily observational review, we discuss substantial progress that now points towards a fundamental connection between early galaxies and reionization. Recent observations with the Hubble Space Telescope

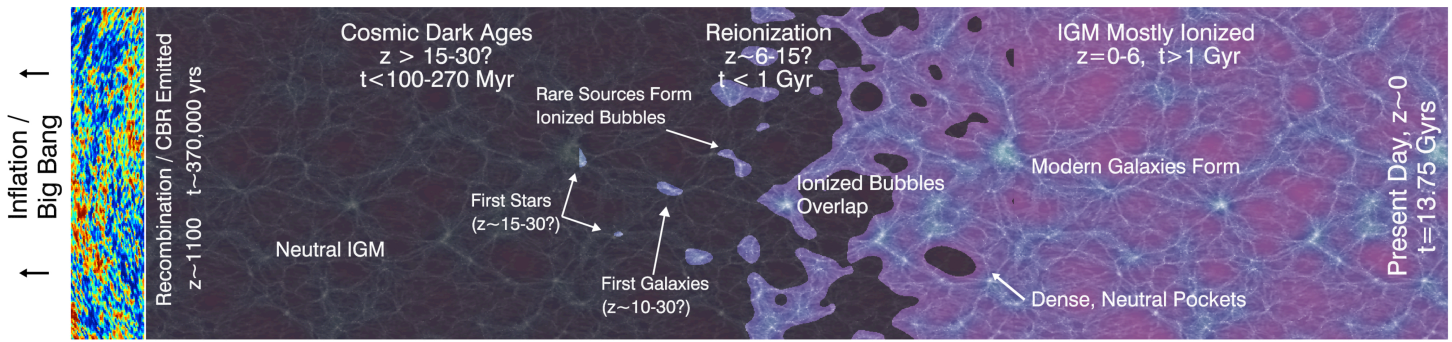

**Figure 1: Cosmic Reionization** The transition from the neutral intergalactic medium (IGM) left after the universe recombined at z~1100 to the fully ionized IGM observed today is termed *cosmic reionization*. After recombination, when the cosmic background radiation (CBR) currently observed in microwaves was released, hydrogen in the IGM remained neutral until the first stars and galaxies[2,4] formed at z~15-30. These primordial systems released energetic ultraviolet photons capable of ionizing local bubbles of hydrogen gas. As the abundance of these early galaxies increased, the bubbles increasingly overlapped and progressively larger volumes became ionized. This reionization process completed at z~6-8, approximately 1 Gyr after the Big Bang. At lower redshifts, the IGM remains highly ionized through radiation provided by star-forming galaxies and the gas accretion onto supermassive black holes that powers quasars.

## BOX 1: The Physics of Reionization

The process of reionization follows the transition from the neutral intergalactic medium (IGM) at high redshift to the ionized IGM we observe locally. The volume ionization fraction of the IGM $Q_{HII}$ progresses from neutral ($Q_{HII}$=0) to fully ionized ($Q_{HII}$=1) according to a changing balance between the production rate of ionizing photons and the rate of recombinations[13-16]. The production rate of IGM ionizing photons,

$$dn_{ion}/dt = f_{esc}\, \zeta_Q\, \rho_{SFR},$$

is the product of the co-moving star formation rate density $\rho_{SFR}$ (in $M_{sun}$ yr$^{-1}$ Mpc$^{-3}$), the number of hydrogen-ionizing photons per second per unit star formation rate $\zeta_Q$ (in s$^{-1}$ $M_{sun}^{-1}$ yr), and the fraction of photons that can escape a galaxy, $f_{esc}$. The rate $dn_{ion}/dt$ is therefore fundamentally tied to the abundance and detailed astrophysics of early stars and galaxies. The recombination rate depends on the IGM temperature and the physical hydrogen density which declines with time according to the universal expansion factor $a^{-3}$, and is enhanced in locally-overdense regions by the "clumping factor" $C_{HII} = <n_H^2>/<n_H>^2$, where $n_H$ is the hydrogen number density. Recent estimates suggest $C_{HII}$~1-6[16,90]. The recombination rate therefore depends primarily on atomic physics and the details of cosmological structure formation.

The observational requirements for determining whether galaxies played the dominant role in causing reionization are summarized as follows. First, the star formation rate density $\rho_{SFR}$ must be quantified, and is typically measured through the observable proxy of the rest frame ultraviolet luminosity density $\rho_{UV}$ via the luminosity distribution of high-redshift star forming galaxies. Second, the number of ionizing photons per unit time $\zeta_Q$ and the escape fraction $f_{esc}$ must be observationally estimated by determining the relative effects of stars, dust, and nebular emission on the rest frame ultraviolet region of high-redshift galaxy spectral energy distributions or by direct observations in the Lyman continuum at lower redshift. To fully reionize the universe, the integrated history of star formation must produce more than one ionizing photon per atom, such that ionizations exceed recombinations. For a standard stellar initial mass function, this requirement can be translated[2] into an effective co-moving stellar mass density of $\rho_{star}$ ~ 2 × $10^6$ $f_{esc}^{-1}$ $M_{sun}$ Mpc$^{-3}$.

agreement on how reionization is expected to unfold (Fig. 1). Sources of ionizing photons, such as star-forming galaxies, are associated with dark matter halos forged through hierarchical structure formation[29]. Initially, these sources are rare and only ionize sparse localised "bubbles" in the dense intergalactic medium. As the Universe expands and the mean IGM density declines, early sources grow in abundance, mass, and luminosity with time and the ionized regions increase in number and extent. The ionized bubbles eventually overlap, allowing the mean free path of ionizing photons and the average volume ionized fraction $Q_{HII}$ in the IGM to increase rapidly. The process is completed with the ionization of neutral pockets of hydrogen gas isolated from luminous sources. A technical description of the reionization process and the key requirements for star forming galaxies to be the responsible agents are provided in **Box 1**.

The complete ionization of hydrogen in the intergalactic medium requires sustained sources of Lyman continuum photons with wavelengths $\lambda$ < 91.2nm. If galaxies are responsible, the process of reionization should mirror their time-dependent density. The number of Lyman-continuum photons produced by star-forming galaxies can be modelled and, assuming some fraction can escape galaxies, the average volume fraction of hydrogen ionized by these photons can be calculated from the star formation rate density. The history of star formation is further constrained by its time integral, the mass density of long-lived stars. At high redshift the presence of such stars is usefully probed by the Spitzer Space Telescope[30,31]. A final constraint is the optical depth to Thomson scattering by free electrons associated with reionization as inferred by their ability to polarise CMB photons. Observational constraints on the electron optical depth[32] suggest that an extended, low-level of ionization (~10% by volume) may be needed to high redshifts (z>20) if reionization gradually completes at z~5.5-8.5. Reconciling the electron scattering optical depth with the number of available ionizing photons inferred from the currently observed star formation rate may therefore be difficult.

The above observations are detailed in **Box 2**. These indirect constraints are broadly consistent with a gradual reionization starting at z~20 and completing at z~6, as first indicated by measures of neutral hydrogen absorption in distant quasar and gamma ray burst spectra that can probe the end of reioinisation[33-36]. Ultimately, we may chart the distribution of ionized bubbles forming within intergalactic neutral hydrogen directly using radio interferometers sensitive to the hyperfine transition at 21cm, a topic not covered explicitly here (for a review see ref. 37). Such technically challenging 21cm observations are still some years away, but with the newly refurbished HST we can already make progress through a direct census of faint star-forming galaxies and studies of their stellar populations.

## A First Census of Early Galaxies

High-redshift galaxies are located by the effect of intervening neutral hydrogen absorption on their colours. Even small amounts of neutral gas can extinguish the light from a galaxy blueward of the Lyman $\alpha$ line in its restframe, causing a "break" in the observed galaxy spectrum. As a result, high-redshift galaxies are often referred to as Lyman Break Galaxies[38-41] (LBGs), and the tell-tale signature of such a distant object is its disappearance or `dropout' in an imaging filter sensitive to the redshifted UV passband where the hydrogen absorption occurs. Identifying the highest redshift galaxies and determining their influence on reionization therefore require an infrared sensitive camera.

Excellent progress has been made using the dropout technique to find high-redshift galaxies since May 2009 following the installation of HST's Wide Field Camera 3, a panoramic imager that includes a powerful infrared capability (WFC3/IR) operating in the wavelength range 850 to 1700nm. Taking into account the infrared field of view, pixel scale and efficiency, this instrument provides a 40-fold improvement in survey speed over the previous generation Near Infrared Camera and Multi-Object Spectrometer (NICMOS) instrument. The HST Wide Field

(HST) have provided the first detailed constraints on abundance and properties of galaxies in the first billion years of cosmic history. With some uncertainties, these data indicate sufficient UV radiation was produced to establish and maintain an ionized universe by redshift z~7, corresponding to ~800 million years after the Big Bang. Further observations of these early systems with current facilities will produce a more robust census and clarify what fraction of the ionizing radiation escaped primitive galaxies. The rapid progress now being made will pave the way for ambitious observations of the earliest known galaxies with upcoming facilities.

## Probes of the Reionization Epoch

Cosmological reionization involves a complex interplay between the strength, distribution, and spectrum of photoionizing sources and the density and spatial structure of intergalactic gas. Theoretical calculations using both analytical methods[13-17] and sophisticated numerical simulations[18-28] to model these complexities have reached broad

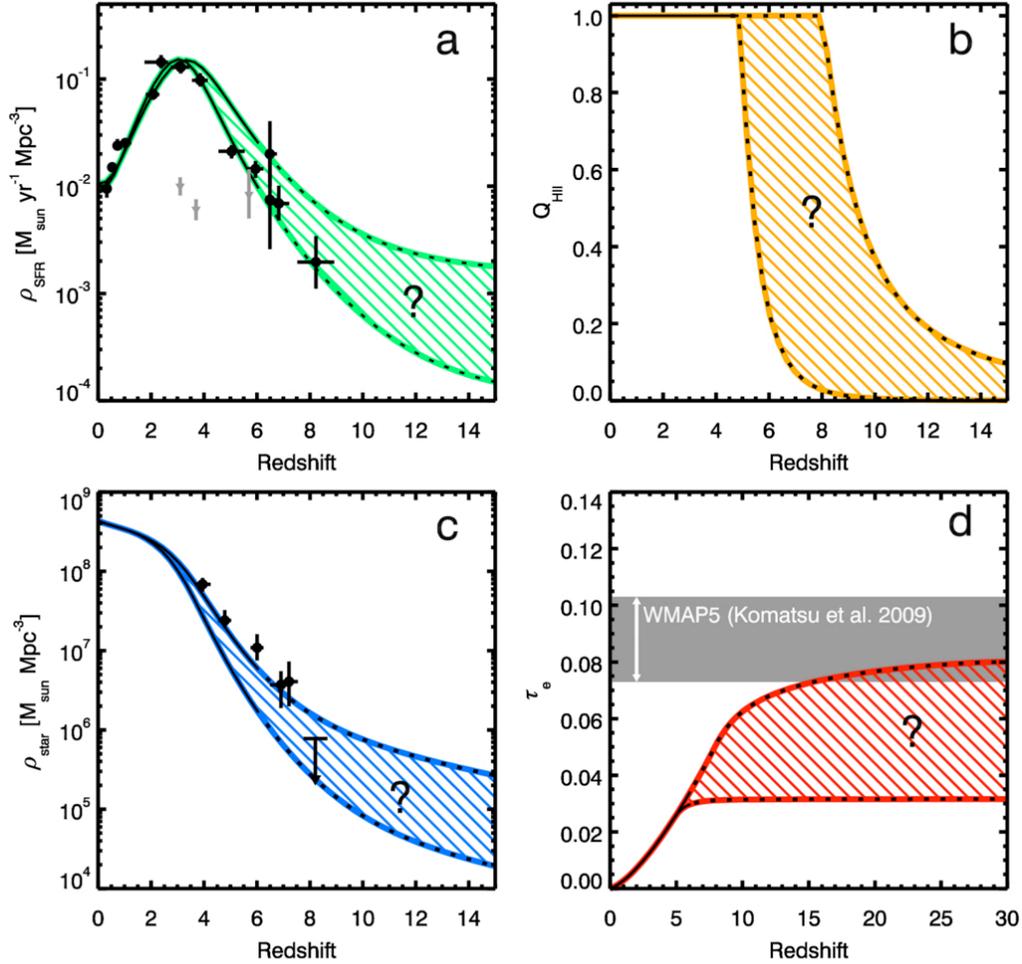

## BOX 2: Observational Probes of the Reionization Epoch

The observed rest-frame UV luminosity density of star forming galaxies, which are the expected photon sources for causing reionization, provides the cosmic star formation rate density $\rho_{SFR}$ (panel **a**, circles with 1 s.d. error bars[40,41,62,91,92], with gray points indicating the contribution from Lyman-α emitting galaxies[65]). Also shown are illustrative model star formation histories for typical stellar ages of $10^8$ yrs consistent with these observations (green area) based on an extension of the fitting form proposed by ref. 93. The models[61,94] span from very metal-poor ($Z\sim5\times10^{-6}\ Z_{Sun}$) stars at the upper boundary to metal-rich ($Z\sim2\ Z_{Sun}$) stars (see details below). The volume fraction of ionized hydrogen $Q_{HII}$ implied by these models is shown in panel **b** (orange region), where, consistent with the present data, the universe becomes fully reionized ($Q_{HII}=1$) at redshifts $z\sim5.5$-$8.5$. The observed stellar mass density (panel **c**, data points, 1 s.d. error bars)[30,31,49-51,95] also constrains the process of reionization since the stellar mass should trace the integral of the star formation rate density (blue shaded area) if most stars are long-lived. Their relative agreement indicates that Population III stars may not contribute significantly to the UV luminosity density at $z\sim7$. Lastly, the scattering optical depth $\tau_e$ of free electrons that polarise the cosmic microwave background (CMB) can also be measured[32] (panel **d**, gray shaded area). The model optical depth $\tau_e$ (red area) can be calculated from $Q_{HII}$ by finding the path length through ionized hydrogen along the line of sight to the CMB. Producing the large electron scattering optical depth given the observed star formation rate density may be difficult without an evolving initial mass function, contribution from Population III stars, or yet unobserved star formation at higher ($z>10$) redshifts.

[We adopt the form $\rho_{SFR}(z) = [a + b(z/c)^h ]/[1 + (z/c)^d ] + g$ with $a = 0.009\ M_{Sun}\ yr^{-1}\ Mpc^{-3}$, $b = 0.27\ M_{Sun}\ yr^{-1}\ Mpc^{-3}$, and $h = 2.5$ . The metal-poor case adopts $c = 3.7$, $d = 7.4$, $g = 10^{-3}\ M_{Sun}\ yr^{-1}\ Mpc^{-3}$, $f_{esc}=0.3$ and $C_{HII}=2$, while the metal rich case uses $c = 3.4$, $d = 8.3$, $g = 10^{-4}\ M_{Sun}\ yr^{-1}\ Mpc^{-3}$, $f_{esc}=0.2$ and $C_{HII}=6$.]

Camera 3 undertook a series of deep images of the Hubble Ultra Deep Field (UDF) – a 4.7 arcmin² area – reaching optimal 5-σ point source sensitivity of ~ 29th magnitude in 3 broad-band filters (see **Box 3**). These data, together with shallower exposures in other areas, have provided the first convincing census of z ~ 7 galaxies and initial indications of galaxy populations at yet higher redshifts[39-47].

The most important achievement from the new WFC3/IR results has been the first robust determination of the volume density of galaxies of different luminosities at z~7 – the luminosity function (LF) - based on over 50 sources seen to date in the various WFC3/IR campaigns. To these HST datasets, one can add constraints based on 22 z~7 candidates similarly detected to brighter limits using the Subaru telescope[48] (see **Box**

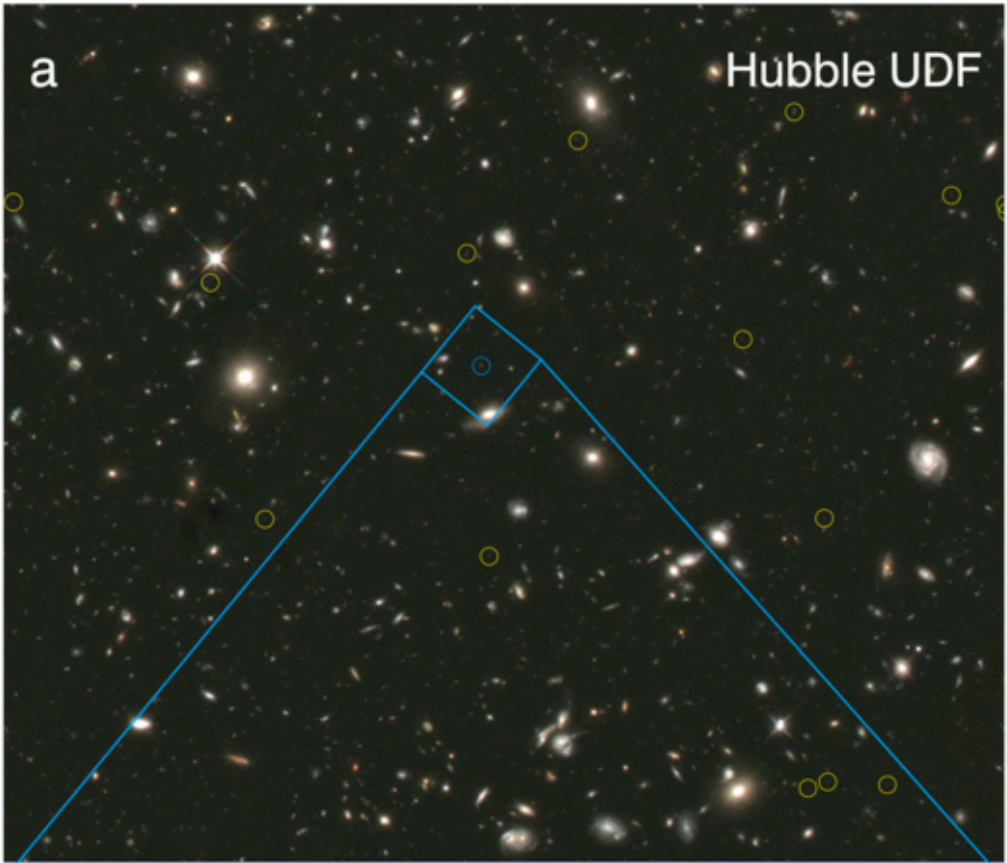

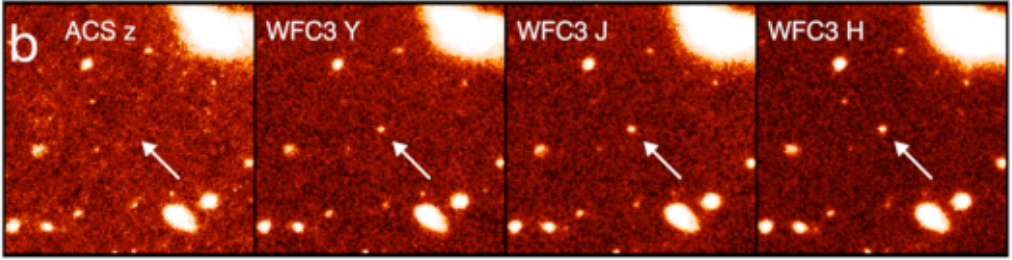

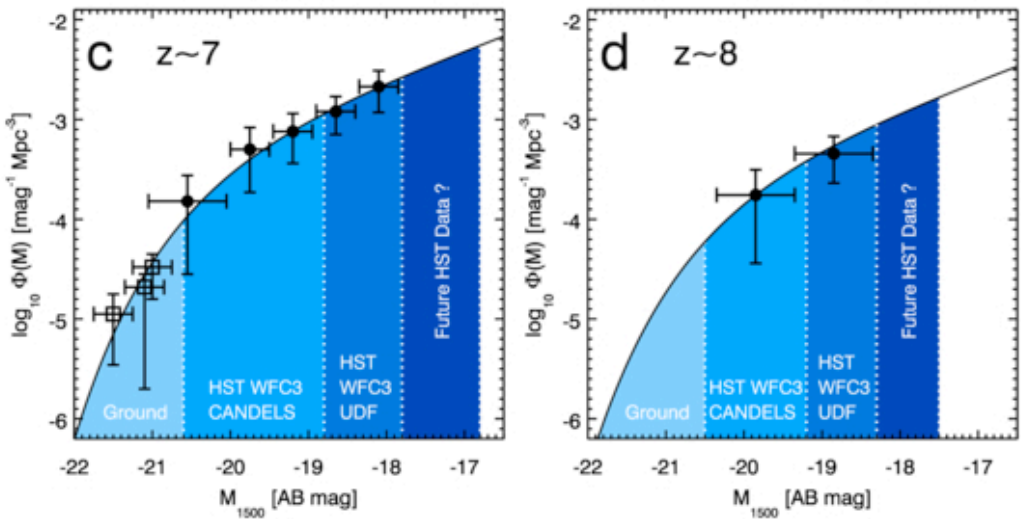

# Box 3: The Discovery and Study of Galaxies at Redshifts z > 7

The search for galaxies at extreme redshifts has been revolutionized by the successful installation of the new camera Wide Field Camera 3 (WFC3) in the Hubble Space Telescope (HST). Panel **a** shows the deepest near-infrared image yet taken with the new WFC3/IR camera of the Hubble Ultra-Deep Field (UDF) with the positions of newly-discovered z ~ 7−8 galaxies[39-41] indicated (circles). For a selected z~7 candidate, panel **b** shows that clear detections in the Y (1.05micron), J (1.2micron) and H (1.6micron) bands with WFC3/IR combined with the non-detection in the optical z-band (0.85micron) with the Advanced Camera for Surveys (ACS) yield a secure photometric redshift[39] of z = 7.2.

The first-year WFC3/IR observations have already produced sufficient numbers of high-redshift galaxies to allow the first reliable determination of the galaxy luminosity function at z = 7 (circles, panel **c**, 1 s.d. error bars), and an initial estimate at z = 8 (circles, panel **d**, 1 s.d. error bars)[39]. Large-area ground-based surveys[48] (open squares, panel **c**, 1 s.d. error bars) are required to uncover the rarest luminous objects at the bright end of the luminosity function, while the newly approved HST Multi-cycle Treasury program, CANDELS, will provide a few hundred galaxies at luminosities around the characteristic luminosity L* at z=7-8. More sensitive imaging in the UDF is both achievable and necessary to measure the number densities of fainter galaxies and to extend these studies to z ~ 9-10.

3), thereby extending the range of the LF. The depth of the Hubble images has been particularly advantageous, as the LF increases steeply for intrinsically fainter sources indicating that the bulk of the UV luminosity density from star-forming galaxies at z~7 emerges from an abundant population of feeble systems.

The most luminous z~7 galaxies have also been detected individually by Spitzer[49,50], and many show strong continua redward of 400 nm in their rest frame indicating established stellar populations of ~$10^{9-10}$ solar masses. Stacking the Spitzer images of the more abundant fainter population gives hints of a marginal signal corresponding to similar stellar populations whose mean mass is $10^{8-9}$ solar masses[51]. The combination of HST and Spitzer has been very effective in addressing the key issue of identifying a sustained source of ionizing radiation. Although uncertainties remain, there is now reasonably convincing evidence that star formation in individual galaxies proceeded at a roughly constant rate over an extended period of 300 million years, corresponding to the redshift range 7<z<10.

## The Escape Fraction of Ionizing Photons

In addition to counting the number of galaxies that produce energetic radiation, we must establish whether a sufficient fraction of the associated UV photons escape to enable reionization. To quantify the production rate of ionizing photons, $dn_{ion}/dt$, and conclusively determine the role of galaxies in cosmic reionization, we thus turn to the problem of determining the number $\zeta_Q$ of energetic Lyman continuum photons per unit star formation rate produced by early stellar populations *and* the fraction $f_{esc}$ of such photons that freely escape a galaxy (see Box 1). Although these quantities almost certainly vary significantly from one galaxy to the next, we can legitimately seek to establish a representative average for the purposes of determining the role of galaxies in reionization.

As UV photons with wavelengths below the Lyman limit (91.2nm) capable of reionizing the intergalactic atoms are rapidly absorbed by neutral gas in the galaxy, the most direct route to estimating $f_{esc}$ would be to measure the emerging flux in this wavelength range. Such observations are intrinsically difficult as typically $f_{esc}\ll1$ and the intervening IGM along the line of sight can absorb the escaping photons, further attenuating the detected Lyman continuum flux. Despite these challenges, intrepid spectroscopic and narrow-band imaging observations[52-55] have detected Lyman continuum flux from galaxies at redshift z~3, the practical redshift limit for this method. These measurements find that the effective escape fraction can vary widely galaxy-to-galaxy, but infer characteristic values of $0.1<\sim f_{esc}<\sim0.2$ at z~3.

The same experiment at redshift z~7 is not technically feasible owing to the increased IGM absorption in high-redshift sources. However, another photometric signature caused by Lyman limit photons as they migrate out of a young galaxy might be observable. If such a photon encounters neutral gas in the galaxy, it will likely ionize a portion of that gas and lead to line emission as well as free-free and bound-free scatterings between electrons and protons. These processes produce nebular radiation whose characteristic emission spectrum can be detected. Models incorporating both the stellar and nebular contributions to galactic emission[56,57] display a spectrum whose power law slope β (where the flux scales with wavelength as $f_\lambda \propto \lambda^\beta$) is strongly connected with the number of escaping ionizing photons through $\zeta_Q$ and $f_{esc}$[42] (see Figure 2). However, unlike the direct measurement of Lyman continuum photons at z~3, this indirect method to estimate $f_{esc}$ from the spectral character of z~7 galaxies has yet to be conclusively demonstrated.

Lower luminosity galaxies in the redshift range 4 < z < 7 show steep UV slopes[58], consistent with the hypothesis that these are relatively-dust free systems[44]. Moreover, the new Hubble data has now indicated that this trend continues to higher redshift where UV slopes with β < -2.5 have been claimed[42,44]. As the youngest starburst galaxies in the local Universe show UV spectra with β > -2.5[58,59], and extreme slopes (β < -3) may indicate Population III stars[60,61], the steep values derived from the new Hubble data are intriguing.

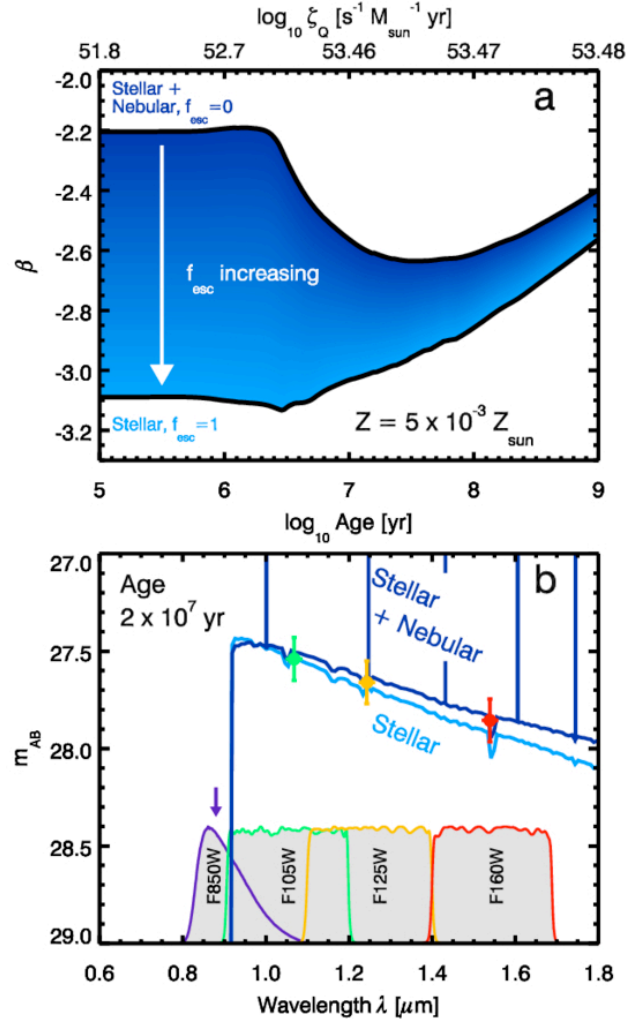

**Figure 2: Ionizing Flux from High-Redshift Galaxies** The co-moving flux of hydrogen ionizing photons $dn_{ion}/dt$ produced by galaxies depends on the total star formation rate density $\rho_{SFR}$, the number of ionizing photons per unit star formation rate $\zeta_Q$, and the fraction $f_{esc}$ of these photons that can escape galaxies to ionize the IGM. Most galaxies at z~7 appear to be nearly dust-free[42], and the escape fraction may therefore reflect the internal ionization rate of gas within each galaxy. This internal ionization produces nebular emission[83] that can redden the spectra of nearly metal-free star-forming galaxies. The colour of the galaxy determined using various filters (panel **b**, shown as shaded areas) may therefore constrain $\zeta_Q$, and $f_{esc}$[57]. Panel **a** shows the UV spectral slope β, defined via the flux density, f (λ) ~ $\lambda^\beta$, for the case of a metal-poor galaxy. We calculate β from stellar population synthesis models[84,85] and our newly-constructed model for the nebular spectrum[83,86-89]. Galaxies with constant star formation rates and $f_{esc}$~1 may appear extremely blue while models with $f_{esc}$~0 are redder owing to nebular emission. Measuring this slope for z~7 galaxies is difficult. Panel **b** shows model high-redshift galaxy spectra with and without nebular emission, along with the synthesized photometry in the available HST filters (data points). The typical UDF photometric uncertainty is ~0.25 magnitudes per source and ~0.11 magnitudes for stacked photometry of 20 objects (1 s.d. error bars). Hence, the current data are insufficiently deep to infer unambiguously $f_{esc}$ and $\zeta_Q$ from the spectral slope.

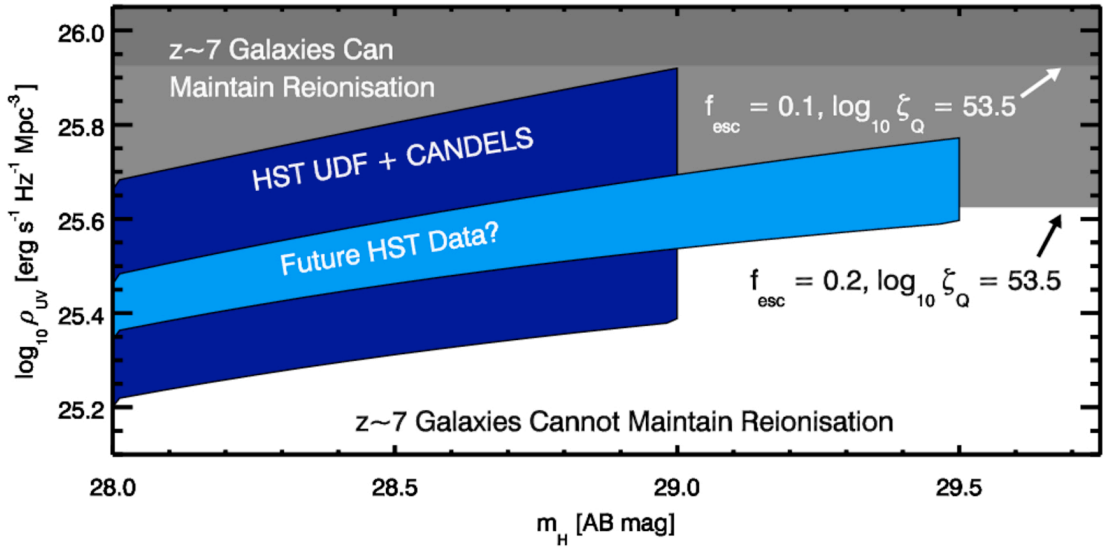

**Figure 3: Distant Star-Forming Galaxies and Reionisation** Expected constraint on $dn_{ion}/dt$ from the combined data of the UDF and the forthcoming CANDELS HST surveys, determined using a Fisher matrix calculation[73]. Shown is the 3-σ uncertainty in the rest-frame ultraviolet luminosity density $\rho_{UV}$ (the observable proxy for the star-formation rate density $\rho_{SFR}$) (dark blue region) as well as the improved constraint achieved by increasing the UDF limiting depth by 0.5 magnitudes (light blue region). The light and dark gray regions show the $\rho_{UV}$ necessary for ionizations to balance recombinations in the IGM for $f_{esc}$=0.2 and $f_{esc}$=0.1, respectively, assuming $\log_{10}\zeta_Q$ =53.5 s$^{-1}$ M$_{sun}^{-1}$ yr (see Figure 2) and $C_{HII}$~2. Increased $C_{HII}$, lower $f_{esc}$, or smaller $\zeta_Q$ require correspondingly larger $\rho_{UV}$ for ionization-recombination balance. The putative escape fraction values are motivated by the steep observed UV slopes of z~7 galaxies[42] and detections of Lyman continuum in z~3 galaxies[52-55]. These estimates assume the measured luminosity function power-law faint-end slope[39] with α=-1.72, and a steeper slope would increase the inferred UV luminosity density and ionizing flux produced by galaxies. While the depth of the UDF and the large area of CANDELS will constrain the abundance of z~7 galaxies well enough to claim consistency with the hypothesis that galaxies trigger reionisation, a definitive test of this question will require deeper data in the UDF with an additional commitment of exposure time comparable to the previous effort. With its new WFC3/IR instrument, HST is capable of performing this exciting experiment in the near future.

The UV slope measurements beyond z~7 remain controversial, partly because of the photometric uncertainties involved. However, if verified with more precise, deeper imaging, current models suggest that such steep slopes most likely indicate nearly dust-free, metal-poor stellar systems with large $\zeta_Q$ or young galaxies with essentially no nebular spectrum, as would be expected if the escape fraction was significant (*e.g.* $f_{esc} > 0.2$). Reliably measuring the UV slopes for a population of z~7-8 galaxies is within reach of HST with sufficiently deep exposures and holds the key to constraining the wanted combination of $\zeta_Q$ and $f_{esc}$.

Our calculations (Figure 3) find that the z~7 photon budget $dn_{ion}/dt$ for reionisation is already balanced with the observed population of galaxies, although with some uncertainty. But importantly, we show that ongoing wider-area HST surveys, as well as future deeper exposures, have the capability of resolving the remaining statistical uncertainties. Finalising the connection between young galaxies and reionisation is thus within reach well before the arrival of impressive new facilities that will study this important reionisation era in greater detail.

## Charting the End of Reionisation

Just as quasars were originally used as distant beacons to probe the intergalactic medium at low redshift, spectroscopy of distant galaxies can test for the presence of neutral hydrogen along the line of sight. A valuable spectral diagnostic in distant star forming galaxies is the Lyman α emission line at rest wavelength of 121.6nm, produced internally by gas heated by young stars. The observability of Lyman α emission is sensitive to the ionization state of the IGM, as the observed line strength can be attenuated by intervening neutral hydrogen. The challenge in utilising this technique to chart the end of reionization lies in isolating the impact of neutral hydrogen from other effects that may diminish the strength of the Lyman α line, including dust.

The abundance of dropout-selected galaxies declines markedly from redshifts z~3 to z~7 [62], yet the fraction of these galaxies showing intense Lyman α emission increases with redshift[63]. An important complementary technique selecting high-redshift Lyman α emitting galaxies (LAEs)[64] through narrow band filters yields a similar result[65] (Fig. 4, panel **a**). As the strong line emitters are generally found to be galaxies with dust-free colours, the increasing fraction of line emitters suggests a reduced obscuration from dust at early times. We therefore expect that most early star-forming galaxies should exhibit prominent Lyman α emission until we reach back to the era when the IGM becomes partially neutral, at which point the fraction of line emitting galaxies should decrease.

Searches for a drop in the fraction of Lyman α emitting galaxies could thus be a very effective probe of when reionization ended[7,9,10]. Much excitement has been generated by a claimed drop in the abundance of Lyman α emitters seen in narrow band imaging surveys with the Subaru telescope[66] in the short time interval (150 million years) between z=5.7 and z=6.6. Further studies of an enlarged area[67] at z=6.6 and a deeper narrowband search[68] at z=7 also indicate a decreasing abundance of Lyman α emitting galaxies at z>6.5. Such a change might highlight an increase in the IGM neutral fraction. The patchy character of the neutral IGM at the end of reionization should also affect the spatial clustering of Lyman α emitters[69], but as no such signal has been detected the question of whether this drop arises from a change in IGM transmission remains unanswered.

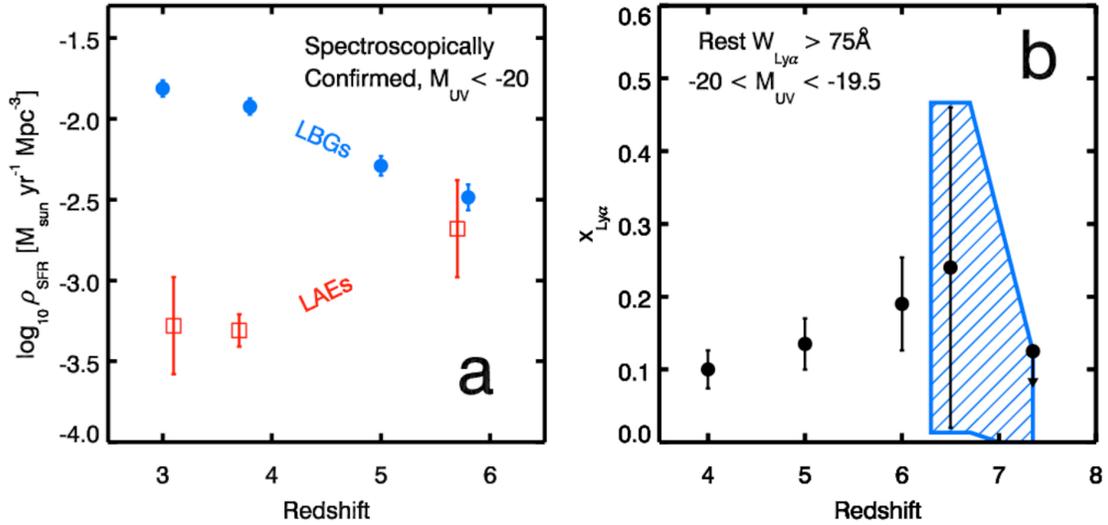

**Figure 4: Tracing the End of Reionization with the Lyman α Line** Recent observations (panel **a**) have revealed an increasing abundance of narrow-band selected Lyman-α emitting galaxies (LAEs, red squares, 1 s.d. error bars)[65] over 3<z<6, when the IGM is known to be highly ionized, while the abundance of colour-selected Lyman-break galaxies (LBGs, blue points, 1 s.d. error bars) declines[62]. In the absence of changes in the IGM ionization state, we therefore expect most dust-free z>6 galaxies to show powerful Lyman α line emission. Observations at z>6 (panel **b**) have instead revealed a possible decline in the prevalence of Lyman-α emitters[66] over 5.7<z<7.0 (black points, 1 s.d. error bars), as expected if reionization ended at z~6. This trend may now be confirmed via spectroscopic surveys of 4<z<7 LBGs which show a tentative decrease in the fraction of Lyman-α emitting galaxies[63,70,71] at z>6.3 (blue hatched region). This result will require confirmation via infrared spectroscopy of z>7 LBGs.

To verify the potentially important decline in the abundance of Lyman α emitters beyond z~6, the fraction of line-emitting galaxies needs to be determined at even earlier times. If the Subaru survey results arise from an increased neutral fraction, we would expect a continued drop in the fraction of line emitting galaxies at higher redshift. A few z~7-8 gravitationally-lensed Lyman break candidates have been found via deep imaging in foreground massive clusters but none has been spectroscopically confirmed[70,71], and thus some candidates may be at lower redshift. However, intriguingly, no Lyman α emission has been seen in any of these lensed sources, suggestive of a continued increase in the neutral fraction (Fig. 4, panel **b**). Recently, there has been a claimed spectroscopic detection of Lyman α line emission from a z=8.6 dropout-selected galaxy[72] found in the new HST data. If confirmed, the detection of a Lyman α line emitter would provide an important milestone in the study of the state of the IGM during the reionization epoch. Extending the Lyman α visibility test with other z>7 candidates newly discovered by HST is therefore an important priority for which results can be expected soon.

## Future Prospects

While the rapid advances in the study of the reionization epoch afforded by the new capabilities of HST have been remarkable, a variety of outstanding observational issues remain. Accurate measures of the abundance of high-redshift galaxies and their luminosity distributions could still need to be improved, and the detailed study of the stellar populations in these galaxies is necessary to robustly predict their output of UV photons. Furthermore, the gaseous content of high-redshift galaxies and the structure of neutral hydrogen in the IGM external to these galaxies are largely unknown. Addressing these outstanding issues will require renewed effort with existing and forthcoming observatories over the next decade.

We argue that the Hubble Space Telescope can further leverage its new infrared capabilities to improve the census of high-redshift galaxies. The current limitations are the precision with which the faint-end slope α of the LF is determined and the possibility of contamination from lower redshift sources. We have carefully evaluated various survey strategies[73] and estimate a now approved larger area multi-cycle HST survey, in combination with the deeper UDF exposures, will improve the bright end of the LF and reduce the uncertainties on α to ±0.15. These observations will provide more precise estimates of the overall UV luminosity and star formation rate densities. Further deep HST imaging via an investment comparable to that undertaken for the current UDF data would improve the constraint on α to ±0.08 (see Figure 3). Such ultradeep exposures, with a deployment of infrared filters carefully arranged to tighten constraints on the UV slopes of z~7 galaxies, can also be used to enhance the recognition of higher redshift galaxies.

Detailed spectroscopic studies of the stellar populations of high-redshift galaxies are required to determine the abundance of young, hot stars, and the importance of the gas and dust in absorbing photons that can reionize the IGM. While comparable sources at lower redshift may soon be within reach of a new generation of infrared spectrographs nearing completion for 8-10 metre class telescopes, the James Webb Space Telescope (JWST)[74] and the next generation of extremely large ground-based telescopes with apertures of 20-40 meters[75-77] will ultimately resolve observational uncertainties about whether strong nebular emission pollutes some of the Spitzer photometry of z~7 galaxies. If so, both the currently derived stellar masses and ages of high-redshift galaxies would be over-estimated[56,57]. Additionally, JWST will deploy infrared cameras and spectrographs designed to probe the rest-frame UV and optical emission from galaxies beyond redshifts z~10.

Forthcoming radio telescopes will inform us to the abundance and distribution of gas fuelling star formation in high-redshift galaxies, as well as the ionization state of hydrogen in the IGM. The Atacama Large Millimeter Array[78] may observe fine-structure carbon and nitrogen lines[79] in distant star-forming galaxies, allowing us to characterize how gas converts into the stars that produce ionizing photons. Future 21cm interferometers include reionization studies as an important science justification[80-82], and will observe the topology of reionization through the spectral signatures of redshifted neutral hydrogen. 21cm observatories will observe neutral hydrogen in the IGM "disappear" as it is ionized during the reionization process, thereby providing an "inverse" experiment to complement rest-frame UV observations with HST and JWST. As these observational facilities come online over the next decade, the reionization epoch will be thoroughly examined.

**Acknowledgements** We thank Anatoly Klypin for the use of his cosmological simulation, and Avi Loeb, Alice Shapley, and Lars Hernquist for useful comments. B.E.R. acknowledges support from a Hubble Fellowship. R.S.E. acknowledges the hospitality of Leiden Observatory. J.S.D. acknowledges the support of the Royal Society via a Wolfson Research Merit award, and also the support of the European Research Council via the award of an Advanced Grant. R.J.M. acknowledges the support of the Royal Society via a University Research Fellowship. D.P.S. acknowledges financial support from an STFC postdoctoral research fellowship and a Schlumberger Research Fellowship at Darwin College.

**Author Contributions** B.E.R. and R.S.E. wrote the main manuscript text. B.E.R. performed the calculations presented in Box 2 and Figures 1-3. J.S.D. and R.J.M. prepared and analysed the data presented in Box 3. D.P.S. prepared and analysed the data presented in Figure 4. All authors reviewed, discussed, and commented on the manuscript.

**Author Information** The authors declare no competing financial interests. Correspondence and requests for materials should be addressed to B.E.R. (e-mail: brant@astro.caltech.edu).